\documentstyle[12pt]{article}
\newcommand{\qmul}{\mbox
{$*$ \hskip -0.80em \raisebox{-0.8ex}{$\scriptstyle \epsilon$}}}
\newcommand{\qodd}{\mbox
{$*$ \hskip -0.75em \raisebox{-0.9ex}{$\scriptscriptstyle 1$}}}
\newcommand{\qeven}{\mbox
{$*$ \hskip -0.75em \raisebox{-0.9ex}{$\scriptscriptstyle 0$}}}
\begin{document}
\renewcommand{\thefootnote}{\fnsymbol{footnote}}
\begin{center}
{\large\bf  Odd Poisson Bracket in Hamilton's Dynamics \\ }
\vspace{1cm}
V.A. Soroka
\footnote{E-mail:  vsoroka@kfti.kharkov.ua}
\vspace{0.25cm}\\
{\it Kharkov Institute of Physics and Technology}\\
{\it 310108, Kharkov, Ukraine}\\
\vspace{1.0cm}
ABSTRACT
\end{center}
\begin{quotation}
{\small\rm Some applications of the odd Poisson bracket to the
description of the classical and quantum dynamics are represented.}
\end{quotation}

\vspace{1.0cm}
\noindent
{\bf 1. Introduction}
\vspace{0.5cm}

Mathematicians ( first of them was Buttin$^1$) proved that in
the phase superspace apart from the usual even Poisson  bracket
there also exists  another bracket of  the Poisson type  namely
the odd Poisson bracket (OB) having the nontrivial Grassmann grading. In
physics  the OB has firstly appeared as an adequate language for  the
quantization   of   the   gauge   theories   in   the  well-known
Batalin-Vilkovisky scheme$^2$.  However, the apparent  dynamical
role of  the OB was not understood  quite well till  papers$^{3,4}$
in which a  possibility of  the reformulation of
Hamilton dynamics on the basis of the OB was proved for the classical
systems having an equal number of pairs of even and odd
(relative to the Grassmann grading) phase coordinates.
Earlier, the prescription$^5$ for the canonical
quantization of the OB was suggested,
and several odd-bracket quantum representations for the canonical
variables were also obtained. In contrast with the even Poisson
bracket case, some of the odd-bracket quantum representations
turned out to be no equivalent$^6$.
Recently the direct connection of the odd-bracket quantum
representations for the canonical variables with the quantization of the
classical Hamilton dynamics based on the OB has been established$^7$.

I concentrate my    attention in
the report on the dynamical aspect of  the OB, that is on
the description with the help of the OB
of both  the classical and quantum dynamics for the systems
in superspace.

The report is organized as follows. The main properties of the odd
bracket are presented in Section 2. In Section 3 it is shown that
Hamilton's equations of motion obtained by means of the even Poisson
bracket with the help of the even Hamiltonian can be reproduced by the odd
bracket using the equivalent odd Hamiltonian.  The odd-bracket quantum
representations for the canonical variables are described in Section 4. In
Section 5 the problem of quantization of the systems with odd bracket is
considered
on the simplest example of the supersymmetric one-dimensional
oscillator.

\vspace{1.0cm}
\noindent
{\bf 2. Properties of the odd Poisson bracket}
\vspace{0.5cm}

First, we recall the necessary properties of various graded
Poisson brackets. The even and odd brackets in terms of the real even
${y_i} = (q^a,p_a)$ and odd $\eta ^i = \theta ^{\alpha }$  canonical
variables have, respectively, the form
\par
$$
\{ A , B \}_o = A \left[\sum_{a = 1}^n
\left(\stackrel{\leftarrow}{\partial}_{q^a}
\stackrel{\rightarrow}{\partial}_{p_a} -
\stackrel{\leftarrow}{\partial}_{p_a}
\stackrel{\rightarrow}{\partial}_{q^a}\right) -
i \sum_{\alpha =1}^{2n}\stackrel{\leftarrow}{\partial}_{\theta ^{\alpha
}} \stackrel{\rightarrow}{\partial}_{\theta^{\alpha }}\right] B\ ;
\eqno {(1)} $$
$$
\{ A , B \}_1 = A \sum^{N}_{ i=1}
\left(\stackrel{\leftarrow}{\partial}_{y_i}
\stackrel{\rightarrow}{\partial}_{\eta ^i} -
\stackrel{\leftarrow}{\partial}_{\eta ^i}
\stackrel{\rightarrow}{\partial}_{y_i}\right) B\ ,
\eqno {(2)}  $$
where $\stackrel{\leftarrow}{\partial}$ and $\stackrel{\rightarrow}{\partial}$
are the right and left derivatives, and the notation $\partial_x =
{\partial \over {\partial x}}$ is introduced.  By introducing apart from
the Grassmann grading $g(A)$ of any quantity $A$ its corresponding bracket
grading $g_{\epsilon }(A) = g(A) + \epsilon \pmod 2$ ($\epsilon  = 0,1$),
the grading and symmetry properties, the Jacobi identities and the
Leibnitz rule are uniformly expressed for the both brackets (1,2) as
\par
$$ g_\epsilon(\{A _{,} B\}_\epsilon) = g_\epsilon(A) + g_\epsilon(B)\pmod 2\ ,
\eqno {(3a)}
$$
$$
\{ A , B \}_\epsilon = -(-1) ^{g_\epsilon (A)g_\epsilon (B)}\
\{ B , A \}_\epsilon\ ,
\eqno {(3b)}
$$
$$
\sum_{(ABC)}(-1) ^{g_\epsilon (A)g_\epsilon (C)}\
\{ A , \{ B , C \}_\epsilon\}_\epsilon = 0\ ,
\eqno {(3c)}
$$
$$
\{ A , B C \}_{\epsilon } = \{ A , B \}_{\epsilon }\ C +
(- 1) ^{g_\epsilon (A)g(B) }\ B \{ A , C \}_{\epsilon }\ ,
\eqno {(3d)}
$$
where (3a)--(3c) have the shape of the Lie superalgebra relations in their
canonical form$^8$ with $g_\epsilon(A)$ being the canonical
grading for the corresponding bracket.

In terms of arbitrary real dynamical variables $x^M = (x^m, x^\alpha) =
x^M(y,\eta)$ with the same number of Grassmann even $x^m$ and odd
$x^\alpha$ coordinates the odd bracket (2) takes the form
\par
$$
\{ A , B \}_1 = A \stackrel{\leftarrow}{\partial}_M \bar \omega^{MN}(x)
\stackrel{\rightarrow}{\partial}_N B\ .
\eqno {(4)}  $$
The matrix $\bar \omega_{MN}$, inverse to $\bar \omega^{MN}$
\par
$$
\bar \omega_{MN}\bar \omega^{NL} = \delta_M^L\ ,
\eqno {(5)}  $$
and consisting of the coefficients of the odd closed 2-form, in view of
the odd bracket properties (3a)-(3c) can be represented in the form of the
grading strength
\par
$$
\bar \omega_{MN} = \partial_M \bar{\cal{A}}_N - (-1)^{g(M)g(N)} \partial_N
\bar {\cal{A}}_M\ ,
\eqno {(6)}  $$
where $g(M) = g(x^M)$ and $\partial_M
= \partial  / \partial x^M$. The coefficients of the 1-form $\bar
{\cal{A}}(d) = dx^M\bar {\cal{A}}_M$ satisfy the conditions
\par $$
g(\bar {\cal{A}}_M) = g(M) + 1\ ,\qquad (\bar {\cal{A}}_M)^+ =
\bar {\cal{A}}_M\  .
\eqno {(7)}  $$
As can be seen from (6) $\bar \omega_{MN}$ is invariant under
gauge transformations
\par $$
\bar {\cal{A}}'_M = \bar {\cal{A}}_M +
\partial_M \bar \chi \eqno {(8)}  $$
with functions $\bar \chi$ as parameters.

\vspace{1.0cm}
\noindent
{\bf 3. Classical dynamics in terms of the odd Poisson bracket}
\vspace{0.5cm}

Let us consider the Hamilton system containing an equal number $n$ of
pairs of even and odd with respect to the Grassmann grading real canonical
variables. We require that the equations of motion of the system be
reproduced both in the even Poisson-Martin bracket (1) with the help of the
even Hamiltonian $H$ and in the odd bracket (2) with the Grassmann-odd
Hamiltonian $\bar H$, that is$^{3,4}$,
\par
$$
{dx^M\over dt} = \{x^M, H\}_0 = \{ x^M, \bar H \}_{1}\ ,
\eqno{(9)}
$$
where t is the proper time. Using definitions (4) and (1) together with
(5),(6) the relations (9) can be represented as the equations \par $$
(\partial_M \bar{\cal{A}}_N - (-1)^{g(M)g(N)} \partial_N \bar{\cal{A}}_M)
\omega^{NL} {\partial_L} H = {\partial_M} \bar H \eqno{(10)} $$ to derive
the unknown $\bar H$ and $\bar{\cal{A}}_M$ under the given $H$ and the
even matrix $\omega^{MN}$ corresponding to the even bracket (1).

In order to solve Eqs. (10) it is convenient to use such real canonical in
the even bracket (1) coordinates $x^M$ which
contain among canonically conjugate pairs the pair consisting of the
proper time $t$ and the Hamiltonian $H$. It follows from Eqs.(9)
that the rest of the canonical quantities $z^{M}$ would be the integrals
of motion for the system considered: even $I_1,...,I_{2(n-1)}$ and
odd $\Theta^1,...,\Theta^{2n}$. In terms of these coordinates $x^{M}$
Eqs. (10) take the form
\par
$$
(\partial_M \bar{\cal{A}}_t - \partial_t \bar{\cal{A}}_M)=
{\partial_M} \bar H\ .
\eqno{(11)}
$$
The quantities $\bar{\cal{A}}_M, \bar \chi$ and $\bar H$ can be
expanded in powers of the Grassmann variables $\Theta^{\alpha }$ as
\par
$$
\bar{\cal{A}}_m = \sum_{k=1}^{n} {i^{(k-1)(2k-1)} \over (2k-1)!}
A_{m\alpha_1\dots \alpha_{2k-1}}\Theta^{\alpha_1 }
\dots\Theta^{\alpha_{2k-1} }\ ,
\eqno{(12a)}
$$
$$
\bar{\cal{A}}_\alpha = \sum_{k=0}^{n} {i^{k(2k+1)} \over (2k)!}
B_{\alpha \alpha_1\dots \alpha_{2k}}\Theta^{\alpha_1 }
\dots\Theta^{\alpha_{2k} }\ ,
\eqno{(12b)}
$$
$$
\bar \chi = \sum_{k=1}^{n} {i^{(k-1)(2k-1)} \over (2k-1)!}
\chi_{\alpha_1\dots \alpha_{2k-1}}\Theta^{\alpha_1 }
\dots\Theta^{\alpha_{2k-1} }\ ,
\eqno{(12c)}
$$
$$
\bar H = \sum_{k=1}^{n} {i^{(k-1)(2k-1)} \over (2k-1)!}
h_{\alpha_1\dots \alpha_{2k-1}}\Theta^{\alpha_1 }
\dots\Theta^{\alpha_{2k-1} }\ .
\eqno{(12d)}
$$
The $\Theta^{\alpha }$ coefficients are the real Grassmann-even functions
of the even variables $x^m = (t,H,I_1,\dots,I_{2(n-1)})$ and are chosen to
be antisymmetric in the indices contracted with $\Theta^{\alpha
}$. In terms of these functions the gauge transformations (8) have the form
\par
$$
A'_{m\alpha_1\dots \alpha_{2k-1}} = A_{m\alpha_1\dots \alpha_{2k-1}} +
\partial_m  \chi_{\alpha_1\dots \alpha_{2k-1}}\ , (k = 1,\dots,n)\ ;
$$
$$
B'_{[\alpha \alpha_1\dots \alpha_{2k}]} = B_{[\alpha \alpha_1\dots
\alpha_{2k}]} + \chi_{\alpha\alpha_1\dots \alpha_{2k}}\ ,
(k = 0,1,\dots,n-1)\ ;
$$
$$
B'_{(\alpha \alpha_j) \alpha_1\dots \alpha_{j-1}\alpha_{j+1}\dots \alpha_{2k}}
= B_{(\alpha \alpha_j) \alpha_1\dots \alpha_{j-1}\alpha_{j+1}\dots
\alpha_{2k}}\ , \left(k = 0,1,\dots,n \atop j = 1,\dots,2k \right)\ ;
\eqno{(13)}
$$
where the expansion in the components with different symmetries of the
indices has been used for the tensor antisymmetric in all indices but the
first
\par
$$
B_{\alpha \alpha_1\dots \alpha_{2k}} = B_{[\alpha \alpha_1\dots
\alpha_{2k}]} + {2 \over {N + 1}} \sum_{j=1}^{N}
B_{(\alpha \alpha_j) \alpha_1\dots \alpha_{j-1}\alpha_{j+1}\dots
\alpha_{2k}}\ .
$$

The additive character of the transformations for the functions
$B_{[\alpha \alpha_1\dots \alpha_{2k}]} (k = 0,1,\dots,n-1)$ allows us to
put them equal to zero in the expression (12b) for $\bar{\cal{A}}_\alpha$
by choosing $\chi_{\alpha\alpha_1\dots \alpha_{2k}} = - B_{[\alpha
\alpha_1\dots \alpha_{2k}]}$. This gauge choice amounts to the following
gauge condition
\par $$
\Theta^{\alpha } \bar{\cal{A}}_\alpha = 0\ .
$$
Using this condition and Eqs.(11), we obtain the equality
\par
$$
\bar H = \bar{\cal{A}}_t\ .
\eqno{(14)}
$$
which, being substituted again into Eqs.(11), leads to the simple equations
\par
$$
{\partial}_t \bar{\cal{A}}_M = 0\ .
\eqno{(15)}
$$
Thus, in consequence of (15), the solution of Eqs.(11) for
$\bar{\cal{A}}_M$ and $\bar H$ in the chosen gauge resides in that the
nonzero coefficients $A_{m\alpha_1\dots \alpha_{2k-1}}$ and $B_{(\alpha
\alpha_j) \alpha_1\dots \alpha_{j-1}\alpha_{j+1}\dots \alpha_{2k}}$ in
expansions (12a,b) for $\bar{\cal{A}}_M$ are the arbitrary functions
(denoted as $a_{m\alpha_1\dots \alpha_{2k-1}a_{m\alpha_1\dots
\alpha_{2k-1}}}$ and $b_{(\alpha \alpha_j) \alpha_1\dots
\alpha_{j-1}\alpha_{j+1}\dots \alpha_{2k}}$, respectively) of all, except
the proper time $t$, even variables  $H$ and $I_1,\dots,I_{2(n-1)}$, and
the odd Hamiltonian is expressed in terms of these functions with the help
of Eq.(14).

Using the gauge transformations (13) with the arbitrary functions
$\chi_{\alpha_1\dots \alpha_{2k-1}}(t,H,I)$, we obtain the general
solution of Eqs.(11) in the arbitrary gauge:
\par
$$
A_{m\alpha_1\dots \alpha_{2k-1}} =
a_{m\alpha_1\dots \alpha_{2k-1}}(H,I) +
\partial_m \chi_{\alpha_1\dots \alpha_{2k-1}}(t,H,I)\ ;
$$
$$
B_{\alpha \alpha_1\dots \alpha_{2k}} =
{2 \over {2k + 1}} \sum_{j=1}^{2k}
b_{(\alpha \alpha_j) \alpha_1\dots \alpha_{j-1}\alpha_{j+1}\dots
\alpha_{2k}}(H,I) +
\chi_{\alpha\alpha_1\dots \alpha_{2k}}(t,H,I)\ ;
$$
$$
h_{\alpha_1\dots \alpha_{2k-1}} =
a_{t\alpha_1\dots \alpha_{2k-1}}(H,I)\ .
$$
Note that the solution of the analogous problem of finding the even
brackets and the corresponding even Hamiltonians, which lead to the same
equations of motion
\par
$$
{dx^M\over dt} = \{x^M, H\}_0 = \{ x^M, \tilde H \}_{\tilde 0}\ ,
$$
has a similar structure but with the difference that the odd quantities
$\bar{\cal{A}}_M, \bar \chi$ and $\bar H$ has to be replaced by the even
ones.

Thus, we extended the notion of the bi-Hamiltonian systems onto the case
when the pairs of the Hamiltonian-bracket, giving the same equations of
motion, have an opposite Grassmann grading.

\vspace{1.0cm}
\noindent
{\bf 4. Quantum representations of the odd Poisson bracket}
\vspace{0.5cm}

The procedure of the odd-bracket canonical quantization
given in$^{5,6}$ resides in splitting all the canonical
variables into two sets, in the division of all the functions
dependent on the canonical variables into classes, and in the
introduction of the quantum multiplication $*$, which is
either the common product or the bracket composition, in dependence on
what the classes the co-factors belong to. Under this, one of the classes
has to contain the normalized wave functions, and the result of
the multiplication $*$ for any quantity on the wave function $\Psi$ must
belong to the class containing $\Psi$. This procedure is the
generalization on the odd bracket case of the canonical quantization rules
for the usual Poisson bracket $\{\dots , \dots \}_{Pois.}$, which, for
example, in the coordinate representation for the canonical variables $q$
and $p$ is defined as \par $$ q\ *\Psi(q) = q \Psi(q)\ ,\qquad p\ *
\Psi(q) = i\hbar\{p, \Psi(q)\}_{Pois.} = - i\hbar
{\partial\Psi\over{\partial q}}\ , $$ where $\Psi(q)$ is the normalized
wave function depending on the coordinate $q$.

In$^{5,6}$ two nonequivalent odd-bracket quantum
representations for the canonical variables were obtained by using two
different ways of the function division. But these ways do not exhaust all
the possibilities. In$^7$ a more general way of the
division is proposed, which contains as the limiting cases the ones given
in$^{5,6}$.

Let us build quantum representations for an arbitrary graded bracket
under its canonical quantization. To this end, all canonical variables are
split into two equal in the number sets, so that none of them should
contain the pairs of canonical conjugates. Note that to make such
a splitting possible for the even bracket (1), the transition has to be
done from the real canonical self-conjugate odd variables to some pairs of
odd variables, which simultaneously are complex and canonical conjugate
to each other.  Composing from the integer degrees of the variables from
the one set (we call it the first set) the monomials of the odd $2s+1$
and even $2s$ uniformity degrees and multiplying them by the arbitrary
functions dependent on the variables from the other (second) set, we thus
divide all the functions of the canonical variables into the classes
designated as $\stackrel{\epsilon}{O}_s$ and $\stackrel{\epsilon}{E}_s$,
respectively.  For instance, in the general case the odd-bracket canonical
variables can be split, so that the first set would contain the even $y_i$
($i = 1,\dots,n\le N$) and odd $\eta ^{n+\alpha}$ ($\alpha = 1,\dots,N-n$)
variables, while the second set would involve the rest variables$^7$. Then
the classes of the functions obtained under this splitting have the form
\par $$ \stackrel{1}{O}_s =
\left(y_i,\eta^{n+\alpha}\right)^{2s+1}f\left(\eta^i,y_{n+\alpha}\right)\ ;
\qquad   \stackrel{1}{E}_s =
\left(y_i,\eta^{n+\alpha}\right)^{2s}f\left(\eta^i,y_{n+\alpha}\right)\ ,
$$
where the factors before the arbitrary function
$f\left(\eta^i,y_{n+\alpha}\right)$ denote the monomials having the
uniformity degrees indicated in the exponents. These classes satisfy the
corresponding bracket relations \par $$
\{\stackrel{\epsilon}{O}_s,\stackrel{\epsilon}{O}_{s'}\}_{\epsilon } =
\stackrel{\epsilon}{O}_{s+s'}\ ;\qquad
\{\stackrel{\epsilon}{O}_s,\stackrel{\epsilon}{E}_{s'}\}_{\epsilon } =
\stackrel{\epsilon}{E}_{s+s'}\ ;\qquad
\{\stackrel{\epsilon}{E}_s,\stackrel{\epsilon}{E}_{s'}\}_{\epsilon } =
\stackrel{\epsilon}{O}_{s+s'-1}\ ,
\eqno{(16)}
$$
and the relations of the ordinary Grassmann multiplication
\par
$$
\stackrel{\epsilon}{O}_s\cdot\stackrel{\epsilon}{O}_{s'} =
\stackrel{\epsilon}{E}_{s+s'+1}\ ;\qquad
\stackrel{\epsilon}{O}_s\cdot\stackrel{\epsilon}{E}_{s'} =
\stackrel{\epsilon}{O}_{s+s'}\ ;\qquad
\stackrel{\epsilon}{E}_s\cdot\stackrel{\epsilon}{E}_{s'} =
\stackrel{\epsilon}{E}_{s+s'}\ .
\eqno{(17)}
$$
It follows from (16),(17), that $\stackrel{\epsilon}{O} =
\{\stackrel{\epsilon}{O}_s\}$ and $\stackrel{\epsilon}{E} =
\{\stackrel{\epsilon}{E}_s\}$ form a superalgebra with respect to the
addition and the quantum multiplication $\qmul$ ($\epsilon = 0,1$) defined
for the corresponding bracket as
\par
$$
\stackrel{\epsilon}{O}'\ \qmul\ \stackrel{\epsilon}{O}'' =
\{\stackrel{\epsilon}{O}',\stackrel{\epsilon}{O}''\}_{\epsilon }\in\
{\stackrel{\epsilon}{O}}\ ;\   \
\stackrel{\epsilon}{O}'\ \qmul\ \stackrel{\epsilon}{E}'' =
\{\stackrel{\epsilon}{O}',\stackrel{\epsilon}{E}''\}_{\epsilon }\in\
{\stackrel{\epsilon}{E}}\ ;\   \
\stackrel{\epsilon}{E}'\ \qmul\ \stackrel{\epsilon}{E}'' =
\stackrel{\epsilon}{E}'\cdot\stackrel{\epsilon}{E}''\in\
{\stackrel{\epsilon}{E}}\ ,
\eqno{(18)}
$$
where $\stackrel{\epsilon}{O}',\stackrel{\epsilon}{O}''\in
{\stackrel{\epsilon}{O}}$  and
$\stackrel{\epsilon}{E}', \stackrel{\epsilon}{E}''\in
{\stackrel{\epsilon}{E}}$. Note, that the classes
$\stackrel{\epsilon}{O}_0$ and $\stackrel{\epsilon}{E}_0$ form
the sub-superalgebra. In terms of the quantum grading $q_{\epsilon }(A)$
of any quantity $A$ \par $$ q_{\epsilon }(A) = \cases{g_{\epsilon
}(A),&for  $A\in{\stackrel{\epsilon}{O}}$;\cr g(A),&for
$A\in{\stackrel{\epsilon}{E}}$,\cr} $$ introduced for the appropriate
bracket, the grading and symmetry properties of the quantum multiplication
$\qmul$ , arising from the corresponding properties for the bracket (3a,b)
and Grassmann composition of any two quantities $A$ and $B$, are uniformly
written as \par $$ q_{\epsilon }(A\ \qmul\ B) = q_{\epsilon }(A) +
q_{\epsilon }(B)\ , \eqno{(19a)} $$ $$ \stackrel{\epsilon}{O}'\ \qmul\
\stackrel{\epsilon}{O}'' = -(-1) ^{q_\epsilon (\stackrel{\epsilon}{O}')
q_\epsilon (\stackrel{\epsilon}{O}'')}
\stackrel{\epsilon}{O}''\ \qmul\ \stackrel{\epsilon}{O}'\ ,
\eqno{(19b)}
$$
$$
\stackrel{\epsilon}{E}'\ \qmul\ \stackrel{\epsilon}{E}'' =
(-1) ^{q_\epsilon (\stackrel{\epsilon}{E}')
q_\epsilon (\stackrel{\epsilon}{E}'')}
\stackrel{\epsilon}{E}''\ \qmul\ \stackrel{\epsilon}{E}'\ ,
\eqno{(19c)}
$$

With the use of the quantum multiplication $\qmul\ $ and the quantum
grading $q_{\epsilon }$ , let us define for any two quantities $A, B$  the
quantum bracket ((anti)commutator) $[A, B \}_{\epsilon }$ (under its
action on the wave function $\Psi$ that is considered to belong to the
class $E$) in the form $^{5-7}$
\par $$
[A, B \}_{\epsilon}\ \qmul\ \Psi  = A\ \qmul\ (B\ \qmul\ \Psi) -(-1)
^{q_\epsilon (A)q_\epsilon (B)} B\ \qmul\ (A\ \qmul\ \Psi)\ .  \eqno{(20)}
$$ If $A, B \in {\stackrel{\epsilon}{E}}$, then, due to (19c), the quantum
bracket between them equals zero. In particular, the wave functions are
(anti)commutative. If $A$ or both of the quantities $A$ and $B$ belong to
the class ${\stackrel{\epsilon}{O}}$, then in the first case,
due to the Leibnitz rule (3d), and in the second one, because of the
Jacobi identities (3c), the relation follows from the definitions (18) and
(20)
\par
$$
[A , B \}_{\epsilon }\ \qmul\ \Psi  = \{A , B\}_{\epsilon }\ \qmul\ \Psi
= (A\ \qmul\ B)\ \qmul\ \Psi\ ,
$$
that establishes the connection between the classical and quantum brackets
of the corresponding Grassmann parity. Note, that the
quantization procedure also admits the reduction to $O_{o}\cup E_{o}$.

The grading $q_\epsilon$ determines the symmetry properties of the quantum
bracket (20). Under above-mentioned splitting of the odd-bracket
canonical variables into two sets, the grading $q_1$ equals unity for the
variables $y_i \in\ \stackrel{1}{O}$, $\eta^i \in\ \stackrel{1}{E}$ ($i =
1,\dots,n\le N$) and equal to zero for the rest canonical variables
$y_{n+\alpha} \in\ \stackrel{1}{E}$, $\eta^{n+\alpha} \in\
\stackrel{1}{O}$ ($\alpha = 1,\dots,N-n$). Therefore, in this case
the quantum odd bracket is represented with the anticommutators between
the quantities $y_i, \eta^i$ and with the commutators for the remaining
relations of the canonical variables.  If the roles of the first and the
second sets of the canonical variables change, then the quantum bracket
is represented with the anticommutators between $y_{n+\alpha},
\eta^{n+\alpha}$ and with the commutators in the other relations.
In$^{5,6}$ the odd-bracket quantum representations were
obtained for the cases $n = 0,N$, containing, respectively, only
commutators or anticommutators.

\vspace{1.0cm}
\noindent
{\bf 5. Quantization of the systems with the odd Poisson bracket}
\vspace{0.5cm}

As the simplest example of using of the odd-bracket quantum
representations under the quantization of the
classical systems based on the odd bracket$^7$, let us consider the
one-dimensional supersymmetric oscillator, whose phase superspace $x^ A$
contains a pair of even $q, p$ and a pair of odd $\eta ^1, \eta ^2$  real
canonical coordinates.  In terms of more suitable complex coordinates $z =
(p - iq)/\sqrt2$, $\eta = (\eta ^1 - i\eta ^2)/ \sqrt2$ and their complex
conjugates $\bar z, \bar \eta$, the even bracket is written as
\par
$$
\{ A, B\}_{0} = iA \left[\stackrel{\leftarrow}{\partial}_{\bar z}
\stackrel{\rightarrow}{\partial}_z -
\stackrel{\leftarrow}{\partial}_z
\stackrel{\rightarrow}{\partial}_{\bar z} -
\left(\stackrel{\leftarrow}{\partial}_{\bar \eta}
\stackrel{\rightarrow}{\partial}_\eta +
\stackrel{\leftarrow}{\partial}_\eta
\stackrel{\rightarrow}{\partial}_{\bar \eta}\right)\right] B
\eqno{(21)}
$$
and the even Hamiltonian $H$, the supercharges $Q_1$, $Q_2$ and
the fermionic charge $F$ have the forms \par $$ H = z\bar z +
{\bar \eta} \eta\ ;\qquad Q_1 = \bar{z} \eta + z \bar\eta\ ; \qquad Q_ 2 =
i(\bar{z} \eta - z \bar\eta)\ ;\qquad F = \eta \bar\eta\ .  \eqno{(22)} $$
The odd Hamiltonian $\bar H$ and the appropriate odd bracket, which
reproduce the same Hamilton equations of motion, as those resulting from
(21) with the even Hamiltonian $H$ (22), i.e., which satisfy the condition
(9), can be taken as $\bar H = Q_1$ and
\par
$$
\{A , B \}_{1} = iA \left(\stackrel{\leftarrow}{\partial}_{\bar z}
\stackrel{\rightarrow}{\partial}_{\eta} -
\stackrel{\leftarrow}{\partial}_{\eta}
\stackrel{\rightarrow}{\partial}_{\bar z} +
\stackrel{\leftarrow}{\partial}_{\bar \eta}
\stackrel{\rightarrow}{\partial}_z -
\stackrel{\leftarrow}{\partial}_z
\stackrel{\rightarrow}{\partial}_{\bar \eta}\right) B\ .
\eqno{(23)}
$$
The complex variables have the advantage over the real
ones, because with their use the splitting of the
canonical variables into two sets $\bar z , \bar \eta$ and $z , \eta $
satisfies simultaneously the requirements necessary for the quantization
both of the brackets (21), (23).  Besides, any of the vector fields
$\stackrel{\epsilon}{X}_{A_i} = - i \{ A_i ,\dots \}_{\epsilon }$ for the
quantities $\{A_i\} = (H, Q_1, Q_2, F)$, describing the dynamics and
the symmetry of the system under consideration, is split into the sum of
two differential operators dependent on either $\bar z, \bar
\eta$ or $z , \eta$ . For instance, from (21)-(23) we have
\par $$
\stackrel{0}{X}_H = \stackrel{1}{X}_{\bar H} = z\partial_z +
\eta\partial_\eta - \bar z\partial_{\bar z} - \bar \eta\partial_{\bar
\eta}\ .  \eqno{(24)} $$
The diagonalization does not take place in terms of the variables
$x^A = (q, p ; \eta^1, \eta^2)$.

In accordance with the above-mentioned splitting of the complex variables,
we can perform one of the two possible divisions all of the functions into
the classes, which are common for both of the brackets (21),(23), playing
a crucial role under their canonical quantization and leading to the same
quantum dynamics for the system under consideration. If
$\bar z, \bar \eta$ are attributed to the first set, then the
corresponding function division is \par $$ \stackrel{\epsilon}{O}_s =
(\bar z \bar\eta)^{2s+1} f (z,\eta)\ ;\qquad \stackrel{\epsilon}{E}_s =
(\bar z \bar\eta)^{2s} f (z,\eta)\ .  $$

If we restrict ourselves to the classes $O_o$  and $E_o$, then
$\Psi \in\ E_o$ and depends only on $z, \eta $  and $A_i \in\ O_o$.
According to the definition (18), the results of the quantum
multiplications $\qodd$ and $\qeven$ of $z , \eta \in\ E_o$ and $\bar z,
\bar\eta \in\ O_o$ on the wave function $\Psi$ are
\par $$
z \ \qodd\ \Psi = z\ \qeven\ \Psi = z \ \cdot\ \Psi\ ;\qquad
\bar\eta\ \qodd\ \Psi =\bar z\ \qeven\ \Psi  = \partial_z\Psi\ ;$$
$$ \ \ \eta\ \qodd\ \Psi = \eta\ \qeven\ \Psi  = \eta\ \cdot\ \Psi\ ;\qquad
\bar z\ \qodd\ \Psi  = - \bar \eta\ \qeven\ \Psi  = \partial_{\eta}\Psi\ .
\eqno{(25)}
$$
The positive definite scalar product of the
wave functions $\Psi_1(z,\eta)$ and $\Psi_2(z,\eta)$ can be determined in
the form$^9$
\par $$
(\Psi_1, \Psi_2) = {1\over \pi} \int \exp
[-(\mid z\mid ^2 + \bar \theta \eta )]\ \Psi_1(z,\eta)\
[\Psi_2(z,\theta)]^+ d\bar\theta\,d\eta\,d(Re z)\,d(Im z)\ ,
\eqno{(26)}
$$
where $\theta$ is the auxiliary complex Grassmann quantity
anticommuting with $\eta$, and the integration over the real and imaginary
components of $z$ is performed in the limits ($-\infty,\infty$). It is
easy to see that with respect to the scalar product (26) the pairs of the
canonical variables, being Hermitian conjugated to each other under the
multiplication $\qodd$, are $z, \bar \eta$ and $\bar z, \eta$, but under
$\qeven$ are $z, \bar z$ and $\eta, -\bar \eta$.

In order to have the action of the Hamiltonian operator, obtained from
the system quantization, on the wave function, we need, as it is
well known, to replace the canonical variables in the classical
Hamiltonian by the respective operators or, which is the same, to
define their action with the help of the corresponding quantum
multiplication $\qmul$. In this connection, in view of (24),(25), we see
that the self-consistent quantum Hamilton operators in the even and odd
cases, being in agreement with the classical expressions (22) for the
equivalent Hamiltonians $H$ and $\bar H$  and giving the same
result at the action on $\Psi(z, \eta) $ , will be respectively
\par
$$
H\ \qeven\ \Psi = z\ \qeven\ (\bar z\ \qeven\ \Psi)\ -
\ \eta\ \qeven\ (\bar \eta\ \qeven\ \Psi)\ ;
\eqno{(27)}
$$
$$
\bar H\ \qodd\ \Psi  = z\ \qodd\ (\bar \eta\ \qodd\ \Psi)\ +
\ \eta\ \qodd\ (\bar z\ \qodd\ \Psi)\ .
\eqno{(28)}
$$
The Hamiltonians (27),(28) are Hermitian relative to the scalar
product (26) and both, due to (25), are reduced to the
Hamilton operator for the one-dimensional supersymmetric oscillator
$H = a^+a + b^+b$ expressed in terms of the creation and annihilation
operators for the bosons $a^{+} = z,\ a = \partial_z$ and
fermions $b^{+} = \eta,\ b = \partial _{\eta }$ respectively, in the
Fock-Bargmann representation (see, for example$^{10}$). The
normalized with respect to (26) eigenfunctions $\Psi_{k,n}(z,\eta)$ of the
Hamiltonians (27),(28), corresponding to energy eigenvalues $E_{k,n} = k
+ n$ ($k = 0,1; n = 0,1,\dots,\infty$) have the form
\par $$
\Psi_{k,n}(z,\eta) = {1\over\sqrt{n!}} \left(\eta\ \qmul\ \right)^{k}\
\left(z\ \qmul\ \right)^{n}1\ .
$$
Note, that another equivalent representation of the quantum supersymmetric
oscillator can be obtained, if the canonical variables $z, \eta$ are
chosen as the first set.
Let us also note that the consideration described above can be extended to
the quantization of a set non-interacting supersymmetric oscillators by
supplement all the canonical variables $z, \bar z, \eta, \bar \eta$ with
the index $i$ $(i = 1,\dots,N)$ over which a summation have to be
performed in the bilinear combinations of the variables in all the
formulas of this section.

Thus, we have demonstrated that
the use of the quantum representations found for the odd bracket$^7$
leads to the self-consistent quantization of the classical Hamilton
systems based on this bracket. We should apparently expect that these
representations are also applicable for the quantization of more
complicated classical systems with the odd bracket.

\vspace{1.0cm}
\noindent
{\bf 6. Acknowledgements}
\vspace{0.5cm}

The author is sincerely thankful to Professor Abdus Salam, the
International Atomic Energy Agency and UNESCO for hospitality at the
International Centre for Theoretical Physics,Trieste. He would also like to
thank the Organizers of the School and Workshop on Variational and Local
Methods in the Study of Hamiltonian Systems, in particular Profs.
A. Ambrosetti, A Bahri, G.F. Dell'Antonio, G. Vidossich, who
kindly give the opportunity to make the report at the Workshop.

This work was supported in part by the Ukrainian State Committee in
Science and Technologies, Grant N 2.3/664, by Grant N UA 6000 from
the International Science Foundation and by Grant N 93-127 from INTAS.

\vspace{1.0cm}
\noindent
{\bf 7. References}
\vspace{0.5cm}

\begin{enumerate}
\item C. Buttin {\it C.r. Acad. Sci. Paris} {\bf 269} (1969) A87.
\item I.A.Batalin, G.A.Vilkovisky, {\it Phys. Lett.} {\bf 102B} (1981) 27.
\item D.V.Volkov, A.I.Pashnev, V.A.Soroka and V.I.Tkach, {\it JETP Lett.} {\bf
44} (1986) 70;
{\it Teor. Mat. Fiz.} {\bf 79} (1989) 117 (in Russian).
\item V.A.Soroka, {\it Lett. Math. Phys.} {\bf 17} (1989) 201.
\item D.V.Volkov, V.A.Soroka, V.I.Tkach, {\it Yad. Fiz.} {\bf 44} (1986) 810
(in Russian).
\item D.V.Volkov, V.A.Soroka, {\it Yad. Fiz.} {\bf 46} (1987) 110 (in Russian).
\item V.A.Soroka, {\it JETP Lett.} {\bf 59} (1994) 219.
\item F.A.Berezin, {\it Introduction in algebra and analysis with anticommuting
variables}
(Moscow State University, 1983) (in Russian).
\item F.A.Berezin, {\it The method of secondary quantization} (Moscow,
Nauka, 1965) (in Russian).
\item A.M.Perelomov, {\it Generalized coherent states and their applications}
(Moscow, Nauka, 1987) (in Russian).

\end{enumerate}
\end{document}